\documentclass[twocolumn,showpacs,amsmath,amssymb,longbibliography,superscriptaddress,showkeys]{revtex4-1}
\usepackage{color,longtable}
\usepackage{graphicx}
\usepackage{amssymb,amsfonts,amsmath}
\usepackage{here}

\begin{document}
\title{Selective 3-dimensional patterning during phase separation of a continuously laminated layer}
\author{Rikuya Ishikawa, Marie Tani and Rei Kurita}

\affiliation{%
Department of Physics, Tokyo Metropolitan University, 1-1 Minamioosawa, Hachiouji-shi, Tokyo 192-0397, Japan
}%
\date{\today}
\begin{abstract}
Control over the physical properties of materials is ubiquitously required in many fields. One means by which this can be achieved is controlling the internal structure of multi-component materials with an eye to enhancing mechanical properties. Here, we focus on self-organized pattern formation in phase separating materials, where microscopic patterns with a smooth, continuous connection may be realized. We propose a feasible method to control pattern formation using phase separation combined with continuous ``lamination'' of material, when material is continuously and homogeneously layered on top of a base. We find that a random droplet pattern, a lamellar pattern, and a cylindrical pattern are formed depending on the lamination rate $V$. We clarify the dynamics of pattern formation, focusing on the mechanism. This study may lead to the creation of new functional materials through artificial pattern control.
\end{abstract}


\maketitle

\section{Introduction}
Compound materials, which contain two or more components, have unique physical properties which differ from the sum of the properties of each component \cite{Smith2000, Pendry2000, Davami2015, Mizuno2016, Tamura2022}. These strongly depend on their internal structure, that is, how the components are arranged. Control over physical properties thus requires explicit design strategies that determine the pattern formed. It is also often necessary that the components be connected smoothly and continuously. In two component systems, such microscopic patterns may be realized spontaneously through phase separation. This may occur when a well-mixed binary system such as a metal alloy or liquid mixture is quenched. As the phase separation proceeds, a bicontinuous pattern is formed when the concentration is symmetric, while a random droplet pattern is formed when the concentration is asymmetric \cite{Cahn1958, Oono1987, Puri1988,Onuki2002, TA2000, Tanaka2005, Hamley, Connor1998, Ball1990, Berthier2001, Golovin2001}. Pattern formation during phase separation is linked to physical properties, leading to the development of new materials. For example, specific patterns formed in high-entropy alloys have attracted much attention for potential applications~\cite{Manzoni2013, Zhang2014}. Pattern formation in collagen gels~\cite{Furusawa2012, Yonemoto2021} is also expected to have applications in tissue engineering\cite{Sionkowska2017, Balakrishnan2011}. 

In the case of homogeneous quenching in bulk, the topology of the pattern formed during phase separation is universally determined solely by the concentrations of each component \cite{Onuki2002}. This has led to previous works focusing on how patterns may be modified using boundary conditions or additives. For example, a layer might be formed near a substrate using wetting \cite{Puri1997, Araki, Shimizu2017}. It has also been shown through numerical simulations that a lamellar pattern may be formed by adding Janus particles \cite{Krekhov2013}. Meanwhile, phase separation by inhomogeneous quenching offers another approach, where the topology of the patterns may be controlled through adjusting how the temperature profile evolves over time. This obviates the need for an additive or boundary condition~\cite{Furukawa1992, Furukawa1994, Liu2000, Krekhov2009, Kurita2017, Tsukada2021, Ishikawa2022}. One such strategy is directional quenching (DQ), in which a quenched region spreads in one direction with constant velocity $V$. It has been reported that three types of pattern are formed depending on $V$ in 2D : random droplets, lamellar, and columnar (or laterally-oriented lamellar) patterns. Recent work has shown that the quenching front behaves as an effective wall that suppresses concentration flux through it, thus selecting for patterns via a finite size effect~\cite{Tsukada2021, Ishikawa2022}. However, despite its benefits, this is hard to realize in three dimensions experimentally. Here, we report that pattern formation similar to DQ may be achieved by lamination or ``coating'' of a mixture onto a pre-existing base at constant speed instead of temperature control. As the thickness evolves, the bounding surface migrates, imitating the movement of a quenching front [see Fig.~\ref{method}]. It is worth noting that lamination can be realized in three dimensions experimentally by coating or deposition. In this study, we perform numerical simulations of three-dimensional phase separation under lamination, investigating how the pattern changes depending on the lamination rate.

\section{Methods}
The dynamics of phase separation under homogeneous quenching is described by the Cahn-Hilliard-Cook (CHC) equation~\cite{Cahn1958, Onuki2002, Cook1970}. Here, we normalize length and time using a correlation length and characteristic time at an initial temperature $T_0$ in an isotropic regime. The normalized CHC equations are given as 
\begin{eqnarray}
\frac{\partial \phi}{\partial t}=\nabla ^2 \lbrack \epsilon \phi + \phi ^3 -\nabla ^2 \phi \rbrack - \nabla \cdot \vec g(\vec r, t),
\label{CH}
\end{eqnarray} 
where $\phi$, $\vec r$ and $t$ are normalized concentration, position, and time, respectively. A normalized temperature $\epsilon$ is defined as $\epsilon = (T-T_c)/(T_0-T_c)$, where $T_0$ and $T_c$ are the initial temperature and the critical temperature, respectively. When $\epsilon > 0$, the mixed state is stable; when $\epsilon < 0$, phase separation occurs. The concentration $\phi$ is normalized by the concentration of each phase after phase separation at $\epsilon = -1$. Assuming local equilibrium and applying the fluctuation-dissipation relation, the correlation function of the dimensionless noise $g(\vec r, t)$ can be described as 
\begin{eqnarray}
\langle \vec g_i (\vec r,t)\vec g_j (\vec {r^{\prime}} , t^{\prime}) \rangle = \Theta \delta_{ij}\delta(\vec r - \vec {r^{\prime}} )\delta (t - t^{\prime}), 
\end{eqnarray} 
where $i, j = x, y, z$ and $\Theta$ corresponds to fluctuation strength. In this simulation, we set $\Theta = 0.001$ for all $\vec r$. We confirm that pattern formation is not sensitive to the amplitude of $\Theta$. The effects of latent heat and hydrodynamic interactions are neglected. We investigate the dynamics of pattern formation for both an asymmetric case, when the mean concentration $\bar \phi = 0.1$, and for a symmetric case, when $\bar \phi = 0$. Here we define regions where $\phi = 1$ as the A phase (white region in Fig.~\ref{state}), and regions where $\phi = -1$ as the B phase (black region in Fig.~\ref{state}).

\begin{figure}[htbp]
\centering
\includegraphics[width=9cm]{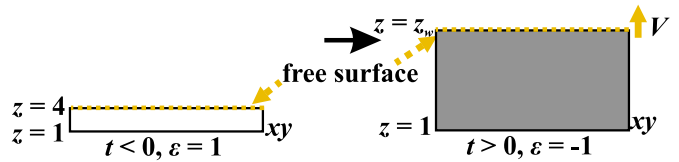}
\caption{Schematic of the simulation method for lamination in the $xz$ plane. Firstly, we anneal a base region at $\epsilon = 1$. The height of the base is 4, so that the system can be regarded as quasi-two dimensional. Next, we quench the system to $\epsilon = -1$ at $t$ = 0. At the same time, the free surface at $z = 4$ starts to move along the z-axis at a constant velocity $V$. The migration of the free surface corresponds to lamination of the surface, or the continuous addition of material. White corresponds to $\epsilon = 1$, while gray corresponds to $\epsilon = -1$. }
\label{method}
\end{figure}

Before lamination, we define a ``base'' region of height 4 with a concentration either equal to the mixture, or set to $\phi$ = $\pm$ 1. The base is equilibrated for a long time at $\epsilon = 1$. A system of this height can be regarded as quasi-two dimensional with respect to phase separating behavior~\cite{Tsukada2019-2}. The system size is then set to $L_x$ : $L_y$ : $L_z$ $=$ 128 : 128 : $z_w(t)$. $z_w(t) = 4 + Vt$, where $V$ is a lamination rate. An upper limit for $z$ is set to $z_w = 128$. We used periodic boundary conditions in the $x$ and $y$ directions and set free surfaces at $z$ = 1 and $z = z_w(t)$ [see Fig.~\ref{method}].

Firstly, we quench the system to $\epsilon = -1$ at $t$ = 0. At the same time, the free surface at $z = 4$ begins to move along the z-axis at speed $V$. The migration of the free surface corresponds to the continuous addition of material i.e. lamination. The dynamics of the phase separation is obtained by solving Equation (\ref{CH}) numerically using the Euler method. Here, we assume that the absorption rate onto the surface during lamination is faster than the phase separation, and that any unevenness of the laminated surface can be neglected since the characteristic length of the phase separation is much longer than the heterogeneity of the surface. All simulations were repeated five times starting from different initial states. 

\section{Results}
\subsection{Pattern formation laminating a $\bar \phi = 0.1$ base}
\begin{figure}[htbp]
\centering
 \includegraphics[width=9cm]{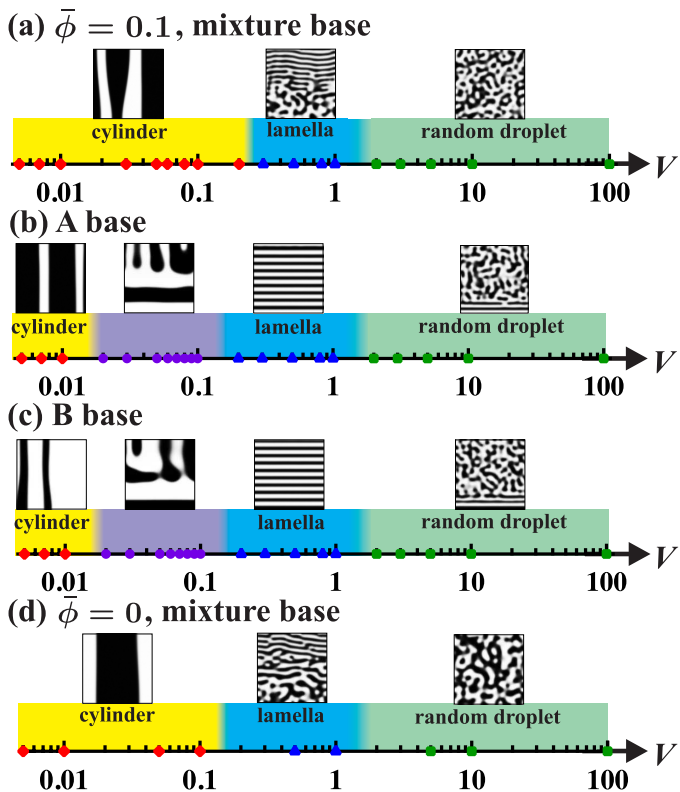}
\caption{$V$ dependence of the pattern formation during phase separation. (a) corresponds to a ``mixture base'' system with $\bar \phi = 0.1$. (b) and (c) correspond to ``A base'' and ``B base'' systems, respectively. (d) corresponds to a ``mixture base'' system with $\bar \phi = 0$. The pattern is classified when the free surface reaches $z$ = 128. Squares, triangles, circles, and diamonds correspond to random droplet, lamellar, lamellae-cylinder coexistence, and cylindrical patterns, respectively.}
\label{state}
\end{figure}

Firstly, we study how pattern formation depends on the lamination rate $V$. We set the mean concentration of the base region to be $\bar \phi = 0.1$. We refer to this base as the ``mixture'' base. Figure \ref{state}(a) shows how the patterns change with respect to $V$. The patterns are classified when the free surface reaches $z$ = 128. Squares, triangles, circles, and diamonds correspond to random droplet, lamellar, lamellae-cylinder coexistence and cylindrical patterns, respectively. 

\begin{figure}[htbp]
\centering
 \includegraphics[width=9cm]{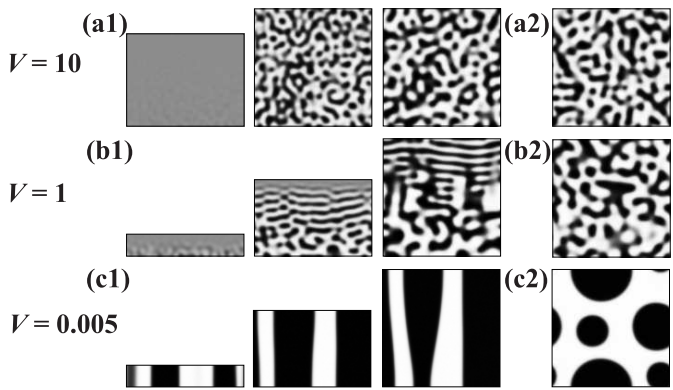}
\caption{Pattern formation in the mixture base system when $\bar \phi = 0.1$. (a1) $xz$ section of the pattern ($y = L_y/2$) when $V = 10$ at $t = 10, 50, 100$ from left to right. (a2) $xy$ section of the pattern ($z = L_z/2$) at $t = 100$. (b1) $xz$ section of the pattern when $V = 1$ at $t = 20, 80, 200$. (b2) $xy$ section of the pattern at $t = 200$. (c1) $xz$ section of the pattern when $V = 0.005$ at $t = 4000, 16000, 40000$. (c2) $xy$ section of the pattern at $t = 40000$. }
\label{pattern}
\end{figure}

When the lamination rate is much faster than the speed of the phase separation, an isotropic droplet phase is formed. Figure \ref{pattern}(a1) and (a2) show the time evolution of the $xz$ section ($y = L_y/2$) of the pattern when $V=10$ at $t = 10, 50$, and 100 from left to right and the $xy$ section of the pattern ($z = L_z/2$) at $t = 100$, respectively. The pattern at high $V$ is almost the same as what is seen under homogeneous quenching.

Meanwhile, the pattern formation is qualitatively changed when the lamination rate is comparable to or less than the speed of the phase separation, that is, $V \le 1$. Figure \ref{pattern}(b1) and (b2) show the time evolution of the $xz$ section ($y = L_y/2$) of the pattern when $V = 1$ at $t = 20, 80$, and 200 and the cross section ($z = L_z/2$) at $t = 200$, respectively. Although a droplet pattern is formed near the bottom, a lamella-like pattern is formed from an intermediate height. However, the lamellar structure is not stable over time, that is, the lamellar structure changes into a droplet pattern during the coarsening process.

Reducing the lamination speed further, the pattern becomes cylindrical at $V \leq 0.2$ (Fig.~\ref{state}(a)). Figure \ref{pattern}(c1) and (c2) show the time evolution of the $xz$ section ($y = L_y/2$) of the pattern when $V = 0.005$ at  $t=4000, 16000$, and 40000, and the cross section ($z = L_z/2$) at $t = 40000$, respectively. Since the lamination is slow, the phase separation occurring at the beginning is quasi-two dimensional i.e. a disk-like pattern is formed in the $xy$ plane. This pattern grows in the $z$ direction as the lamination proceeds, establishing a cylindrical pattern. 

\begin{figure}[htbp]
\centering
 \includegraphics[width=9cm]{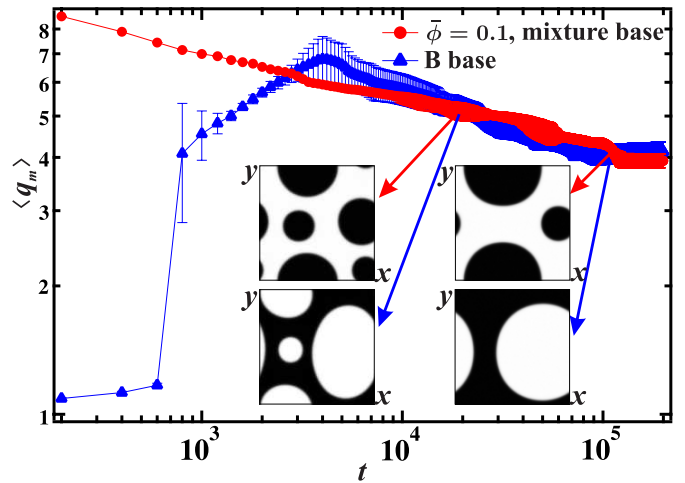}
\caption{Time evolution of $\langle q_m (t) \rangle$ at $V$ = 0.005. Circles and triangles correspond to $\langle q_m (t) \rangle$ on laminating the mixture base system and the B base system, respectively. $\langle q_m (t) \rangle$ decrease over time at later times, indicating that the cylinders are getting thicker in both systems. It is noted that $\langle q_m (t) \rangle$ increases at an early stage in the B base system. Spinodal decomposition starts at $t \sim 600$, followed by the emergence of new cylinders. }
\label{coarsening}
\end{figure}

We also investigate the time evolution of the cylindrical pattern. To characterize the cylindrical pattern, we define a mean wavenumber $\langle q_m \rangle$ which corresponds to the mean diameter of the cylindrical pattern in the $xy$ plane as follows : 
\begin{eqnarray}
q_m (z, t) &=& \frac{\int dq \ q S (q, z, t)}{\int dq \ S (q, z, t)} \\ 
\langle q_m (t) \rangle &=&\frac{1}{z_w}  \int _1^{z_w}dz \ q_m (z, t), \label{qm}
\end{eqnarray} 
where $S (q, z, t)$ is a structure factor obtained by Fourier transforming the pattern in the $xy$ plane at $z$ and $t$. We confirm that $\langle q_m (t) \rangle$ is almost same as the peak position of $S (q, z, t)$. Figure~\ref{coarsening} shows the time evolution of $\langle q_m (t) \rangle$ (circle symbol). The error bar corresponds to the standard deviation of $q_m (z, t)$ in the $z$ direction. We find that $\langle q_m (t) \rangle$ decreases over time, indicating that the cylinders are getting thicker. This means that the cylindrical pattern is stable. Here, the time evolution of $\langle q_m (t) \rangle$ in this system ($t^{-0.1}$) is much slower than that for a droplet pattern in two dimensions ($t^{-1/3}$)~\cite{Onuki2002}. This may be significant, but the system is not large enough to investigate the coarsening dynamics. Larger systems should be investigated in the future.

\subsection{Pattern formation laminating an A or B base}
\begin{figure}[htbp]
\centering
 \includegraphics[width=9cm]{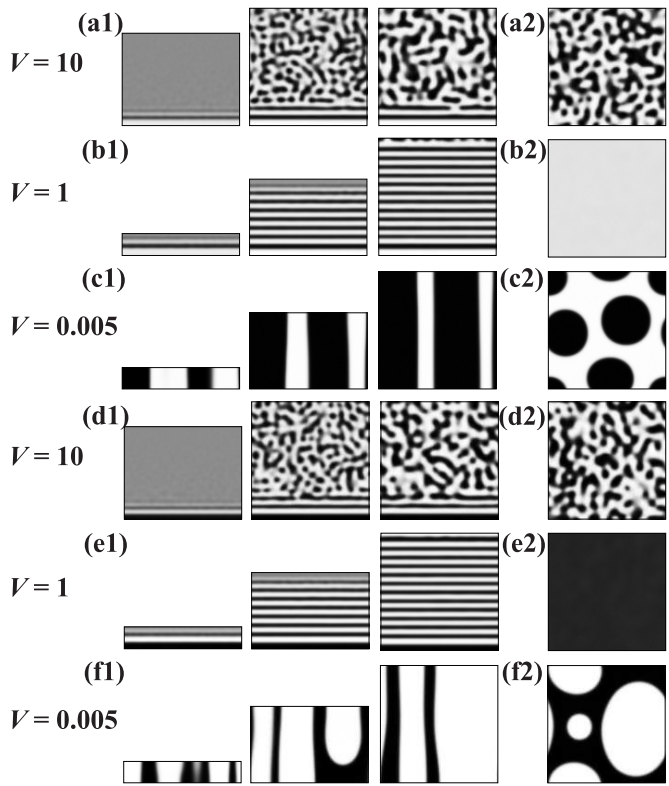}
\caption{Pattern formation on (a)-(c) an A base and (d-f) a B base. (a1) and (d1) show the time evolution of the $xz$ section of the pattern when $V = 10$ ($y = L_y/2$) at $t = 10, 50, 100$, from left to right. (a2) and (d2) show the $xy$ section of the pattern ($z = L_z/2$) at $t = 100$. (b1) and (e1) show the time evolution of the $xz$ section of the pattern when $V = 1$ at $t = 20, 80, 200$. (b2) $xy$ section of the pattern at $t = 200$. (c1) and (f1) show the time evolution of the $xz$ section of the pattern when $V = 0.005$ at $t = 4000, 16000, 40000$. (c2) and (f2) show the $xy$ section of the pattern at $t = 40000$. }
\label{ABbasepattern}
\end{figure}

In the mixture base system, it was found that the pattern growth in the newly laminated region strongly depends on the pattern formed at the base. Here, we investigate the base dependence of the pattern formation further by laminating the same mixture ($\bar \phi = 0.1$) on a $\bar \phi = 1$ base (A base) or a $\bar \phi = -1$ base (B base). Figure \ref{state}(b) and (c) show the patterns formed at different $V$ on the A base and B base, respectively; Figure \ref{ABbasepattern} also shows the time evolution of the pattern on the A base (a)-(c) and the B base (d)-(f). For $V \ge 2$, a random droplet pattern is formed since the lamination is faster than the phase transition, although lamellae are formed near the bottom (Fig.~\ref{ABbasepattern}(a) and (d)). This is similar to pattern formation on a wettable surface~\cite{Araki} and clearly indicates that the base plays the role of a wetting surface. 

\begin{figure}[htbp]
\centering
 \includegraphics[width=9cm]{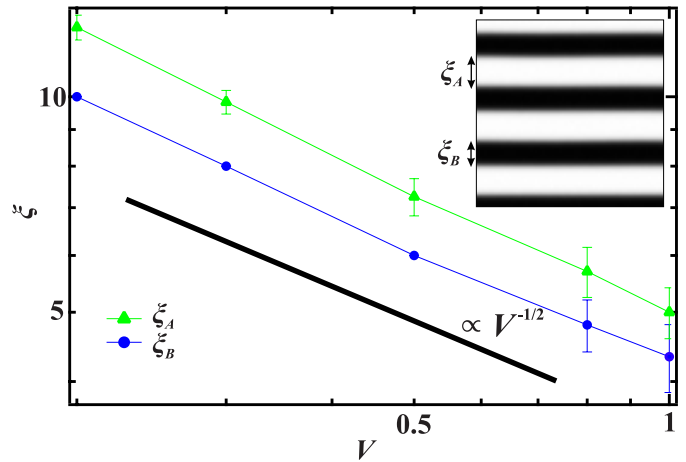}
\caption{$V$ dependence of $\xi_A$ and $\xi_B$ in the B base system. Triangles and circles correspond to $\xi_A$ and $\xi_B$, respectively. $\xi_A$ and $\xi_B$ correspond to the thickness of layers formed by the A and B phases, respectively. Both $\xi_A$ and $\xi_B$ are proportional to $V^{-1/2}$ (solid lines). }
\label{lamella}
\end{figure}

A lamellar pattern is formed over the whole system when $0.2 \le V \le 1$  (Fig.~\ref{ABbasepattern}(b) and (e)). Compared to laminating on a mixture base (see Fig.~\ref{pattern}(b)), the lamellar pattern is rigid with flat surfaces. Here, we note that the lamellar pattern formed on an A or B base system is stable over time. We investigate the layer thickness of the lamellar pattern $\xi$ in both systems. We define the thickness of A phase layers and B phase layers as $\xi_A$ and $\xi_B$, respectively. Figure \ref{lamella} shows the $V$ dependence of $\xi_{A}$ and $\xi_{B}$. Symbols show the mean of $\xi$ while the error bars correspond to the standard deviation over multiple layers. Note that the thickness of layers near the upper and lower boundaries of the system are omitted from the averaging. It is natural that $\xi_A$ is larger than $\xi_B$ since $\bar{\phi} = 0.1$. Here, it is found that $\xi_A$ and $\xi_B$ are proportional to $V^{-1/2}$, as shown by the solid lines in Fig.~\ref{lamella}. This is because thicknesses are determined by competition between diffusion and lamination. The time scale $\tau$ over which a component diffuses across layer may be written $\tau \sim \xi^2/D$, where $D$ is a diffusion constant. When $V > 1/\tau$, the diffusion is slower than the lamination, and the layer cannot be broader, simply because the patterning cannot correlate over longer length scales. Thus, a relation $V = 1/\tau = D/\xi^2$ is derived, giving $\xi \sim V^{-1/2}$. 

\begin{figure}[htbp]
\centering
 \includegraphics[width=9cm]{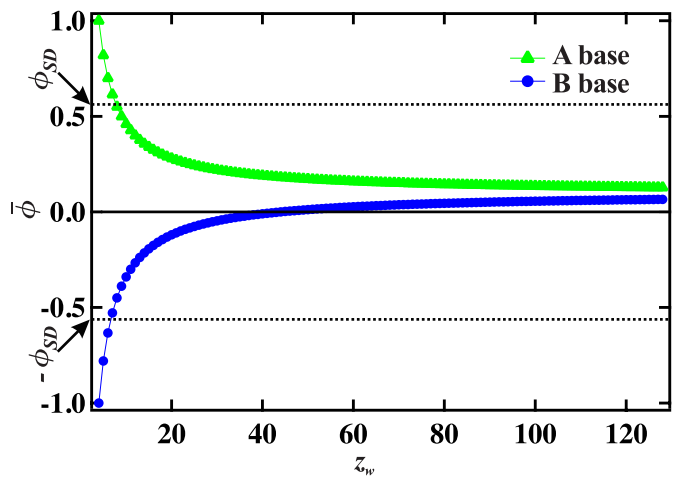}
\caption{Sample height $z_w$ dependence of the mean concentration of the whole system $\bar \phi$ in the B base system. $\bar \phi$ gradually changes as a mixture with $\phi =0.1$ is laminated. When the base is the B phase, a time regime exists where $\bar \phi \le 0$. In this regime, the B phase is the majority. Meanwhile, the A phase is always the majority if the base is the A phase. The dotted lines correspond to $\phi_{SD}$ or $-\phi_{SD}$. 
}
\label{tempora}
\end{figure}

When $V \le 0.1$, the lamellar pattern becomes unstable during the lamination (Fig.~\ref{ABbasepattern}(c) and (f)). This instability comes from the emergence of fluctuations at the interface, as previously reported in Ref.~\cite{Ishikawa2022}. When $V \le 0.01$, a cylindrical pattern emerges from the bottom. Just like in the mixture base system, phase separation proceeds in the bottom quasi-two dimensional slab during the lamination. Here, it is interesting that the cylinder phase is reversed, that is, the cylinder is the B phase on the A base system, while it is the A phase on the B base system (Fig.~\ref{ABbasepattern}(c2) and (f2)). Since the A phase is the majority phase, the interface area becomes larger when the A phase forms cylinders. Thus, the pattern shown in Fig.~\ref{ABbasepattern}(f) would be higher in terms of the interfacial energy. However, at an early stage, when the lamination has not fully progressed, $\bar{\phi}$ averaged over the whole system strongly depends on the base. Figure \ref{tempora} shows the sample height $z_w$ dependence of $\bar \phi$ in the A base system and the B base system. When the base is composed of the B phase, $\bar{\phi}$ is temporarily negative at the beginning. Thus, the A phase is temporarily the minority phase, causing cylinders of A phase to be formed when $z_w$ is small. These cylinders go on to grow in the $z$ direction as the lamination proceeds, even though the A phase becomes the majority. We also compute the time evolution of $\langle q_m \rangle$ in the B base system, shown as triangles in Fig.~\ref{coarsening}. Before $t \sim$ 600 or $z_w$ = 7, the system is metastable since $\bar \phi < - \phi_{SD} (\phi_{SD} = \sqrt{3}/3)$, where $\phi_{SD}$ is a spinodal concentration, shown as a dotted line in Fig.~\ref{tempora}. Phase separation does not occur in this short period. As time progresses, $\langle q_m \rangle$ increases for $600 \le t \le 3600$ or $7 \le z_w \le 22$. In this regime, $\bar{\phi}$ also increases over time and spinodal decomposition starts to occur. It was also found that new small cylinders emerge during this time. After $t > 3600$, these cylinders only coarsen. Similar to the mixture base system, it was found that the cylinders are stable. Meanwhile, the A phase is always the majority when the base is the A phase (Fig.~\ref{tempora}). Thus, the cylinder is formed by the B phase. 

\subsection{Pattern formation laminating a $\bar{\phi} = 0$ base}
\begin{figure}[htbp]
\centering
 \includegraphics[width=9cm]{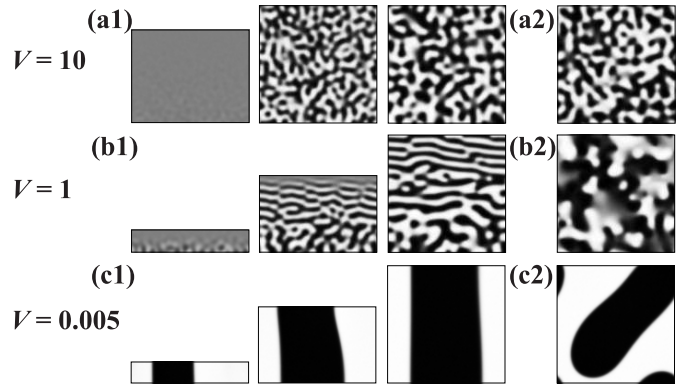}
\caption{Pattern formation in the mixture base system when $\bar \phi = 0$. (a1) Time evolution of the $xz$ section of the pattern when $V = 10$ ($y = L_y/2$) at $t = 10, 50, 100$ from left to right. (a2) $xy$ section of the pattern ($z = L_z/2$) at $t = 100$. (b1) Time evolution of the $xz$ section of the pattern when $V = 1$ at $t = 20, 80, 200$. (b2) $xy$ section of the pattern at $t = 200$. (c1) Time evolution of the $xz$ section of the pattern when $V = 0.005$ at $t = 4000, 16000, 40000$. (c2) $xy$ section of the pattern at $t = 40000$.}
\label{phi0ordinarypattern}
\end{figure}

Finally, we investigate the pattern formation when the base is $\bar{\phi} = 0$. It is well known that the pattern becomes bicontinuous at $\bar{\phi} = 0$, while it is a random droplet pattern for $\bar{\phi} \neq 0$~\cite{Onuki2002}. This difference is also apparent in our mixture base lamination system. Figure \ref{state}(d) shows the pattern as a function of $V$, while Fig.~\ref{phi0ordinarypattern} shows the pattern formation dynamics. When $V \ge 2$, a three-dimensional bicontinuous pattern is formed since the lamination is too fast and the system is almost the same as a homogeneously quenched system. When $V \sim 1$, an unstable lamella pattern is formed similar to the lamella pattern at $\bar{\phi} = 0.1$. For even lower $V$, a pillar is formed with a cross section which is not circular but peanut shaped. At the beginning, a two-dimensional bicontinuous phase is formed in the $xy$ plane. Then, the pattern coarsens over time in $xy$ plane while simultaneously propagating into the $z$ direction. One of the two phases is subsequently isolated rather than maintaining a bicontinuous structure. We confirm that the B phase is incidentally isolated in Fig.~\ref{phi0ordinarypattern}(c2), but the A phase is isolated in a different simulation run. There are two reasons for this: a bicontinuous pattern cannot be realized geometrically in a two-dimensional plane, and the system size is too small.

We also note that the pattern formation on the $\phi = 1$ base is the same as the A base system when $\bar{\phi} = 0.1$. $\bar{\phi}$ is always positive (or negative) since the concentration of the mixture is $\phi = 0$. The pattern formation is determined at the beginning when the system is almost the same as the A base system.

\section{Discussion}
Here, we discuss the relevance of this mode of pattern formation to three-dimensional directional quenching (3D-DQ)~\cite{Ishikawa2022}. In 3D-DQ, a layer of A phase is always formed at the bottom when $\bar{\phi} = 0.1$, stabilizing a lamellar structure. Thus, the mixture base case is fundamentally different from 3D-DQ, although the $V$ dependence of the pattern with lamination seems similar. One of the most important reasons for this is the phase separation dynamics at the beginning. The quenching front in 3D-DQ acts as an effective wall which suppresses flux, whereas the lamination surface in this study completely separates the space. Thus, the phase separation occurs two dimensionally in the lamination system, while a layer of A phase is formed in 3D-DQ. This difference at the beginning affects the pattern growth at later stages. Meanwhile, the $V$ dependence of the pattern in the lamination system with an A or B base is close to that in the 3D-DQ since the base is already established. For example, a lamellar structure is stable over time. However, for small $V$, phase separation occurs two dimensionally after all and leads to a cylindrical pattern; this is never seen in 3D-DQ~\cite{Ishikawa2022}. We conclude that flux beyond the front is crucial for pattern formation in inhomogeneous quenching systems.

\section{Summary}
In this paper, we investigated self-organized pattern formation with lamination. This method is more feasible for experimentally controlling internal patterns using phase separation, in clear contrast to 3D-DQ. It is found that a random droplet pattern, a lamellar pattern, and a cylindrical pattern are formed depending on the lamination rate $V$. The lamellar pattern formed in the mixture base system ($\bar{\phi}$ of the base is the same as that of the laminated mixture) is unstable over time, but it is stable when the base is composed of the A or B phase. The width of the lamellae can be controlled by adjusting $V$. It is also found that the three-dimensional pattern is strongly influenced by the pattern formed at the beginning. For example, phase separation may initially occur in a confined two-dimensional space, followed by propagation of the pattern in the $z$ direction. This mechanism is qualitatively different from phase separation in bulk, and may lead to the creation of new functional materials via rational strategies to achieve pattern control. 

M. T. was supported by JSPS KAKENHI Grant Number 20K14431. R. K. was supported by JSPS KAKENHI Grant Number 20H01874.

\bibliography{base.bib}

\end{document}